\pdfoutput=1
\pdfoutput=1
\pdfoutput=1
\pdfoutput=1
\pdfoutput=1
\documentclass[groupedaddress,nofootinbib,eqsecnum]{revtex4}

\usepackage[dvips]{color}
\usepackage{epsfig}
\usepackage{amsmath}
\usepackage{graphicx}

\begin{document}

\title{Cosmological constraints on Agegraphic dark energy in DGP braneworld gravity}

\author{H. Farajollahi}
\email{hosseinf@guilan.ac.ir}
\affiliation{Department of Physics, University of Guilan, Rasht, Iran}
\author{A. Ravanpak}
\email{a.ravanpak@vru.ac.ir}
\affiliation{Department of Physics, Vali-e-Asr University, Rafsanjan, Iran}
\author{G. F. Fadakar}
\email{gfadakar@guilan.ac.ir}
\affiliation{Department of Physics, University of Guilan, Rasht, Iran}

\begin{abstract}
A proposal to study the original and new agegraphic dark energy in DGP braneworld cosmology is presented in this work. To verify our model with the observational data, the model is constrained by a variety of independent measurements such as Hubble parameter, cosmic microwave
background anisotropies, and baryon acoustic oscillation peaks. The best fitting procedure shows the effectiveness of agegraphic parameter $n$ in distinguishing between the original and new agegraphic dark energy scenarios and subsequent cosmological findings. In particular, the result shows that in both scenarios, our universe enters an agegraphic dark energy dominated phase.
\end{abstract}

\keywords{Agegraphic dark energy; DGP; Brane cosmology;}

\maketitle
\newpage

\section{Introduction}

The results from recent cosmological
observations in cosmic microwave
background (CMB) anisotropies \cite{Spergel}\cite{Larson}, Hubble parameter (Hub) \cite{Xu}, redshift-distance relationship,
of type Ia supernovae (SNIa)  \cite{Riess}\cite{Kowalski}, baryon
acoustic oscillation (BAO) peaks \cite{Frieman}, gamma-ray bursts (GRBs) \cite{Vitagliano} and linear growth of large-scale
structures (LSS) \cite{Tegmark}\cite{Seljak} strongly verifies the universe expansion. These probes also indicate that the current universe due to some kind of negative pressure form of matter dubbed as dark energy is accelerating \cite{Peebles}\cite{Padmanabhan}.

In general, most of the dark energy proposals given in last decay are either not able to explain all features of the universe or introduce too many free parameters have to be fitted with the observational data \cite{Copeland}. Alternatively many theoretical attempts intend to shed light on
the problem in the framework of more fundamental theories such as string theory or quantum gravity.
Although, a complete and comprehensive formulation of quantum gravity has not yet been established, efforts have been made to discover the nature of dark energy according to some principles of the theory. In particular, the so-called holographic dark energy (HDE) models are to be credited as an alternative candidate for dark energy \cite{Horava}\cite{Wei}. They are consistent with
quantum principle, in the sense that they obey the Heisenberg type uncertainty
relation, and predict a time-varying dark energy equation of state (EoS). The widely studied HDE model is very successful in explaining the observational data. In HDE models by choosing the event horizon of the
universe as the length scale, the universe accelerating expansion phase is obtained naturally without resorting to a cosmological constant. However, there
is an obvious drawback in holographic approach with regards to causality appears in the theory. Event horizon is a global concept of space time and determines by future evolution of the universe. It only exists for universe with forever accelerated expansion  \cite{Guberina}\cite{Wei1}.

Alternatively, a new dark energy model, dubbed as
agegraphic dark energy (ADE) model, proposed by Cai, is considered as a solution to the causality problem in the holographic dark energy. In agegraphic dark energy models both uncertainty relation of quantum mechanics and gravitational effect in general
relativity are taken into account \cite{Cai}\cite{Wei2}. In the so called original ADE model, the age of the universe is chosen as the length measure,
instead of the horizon distance and therefore the causality problem is resolved. These models, where assume that the observed dark energy originated from spacetime
and matter field fluctuations in the universe, have been constrained by
various astronomical observation. In fact,  from quantum fluctuations of
spacetime, the Karolyhazy relation \cite{Karolyhazy}\cite{Karolyhazy1} states that the distance $t$ in Minkowski spacetime cannot be
known to a better accuracy than $\delta t=\beta t_p^{2/3}t^{1/3}$
where $\beta$ is a dimensionless constant of order unity. This relation together with the time-energy uncertainty
 relation in quantum mechanics lead to a energy density of quantum
fluctuation of Minkowski spacetime metric given by \cite{Maziashvili}
\begin{eqnarray}
\rho_{D}&\sim&\frac{1}{t_p^2t^2}\sim\frac{M_p^2}{t^2}\label{rho}
\end{eqnarray}
where $t_p$ and $M_p$ are the reduced Planck time and mass, respectively. Therefore, in ADE proposal the energy density, which resembles the HDE, exists within a causal patch and obey the holographic entropy bound.

Independent of the above subject, in recent years,
extra dimensional theories, in which the universe is realized as a brane embedded in a higher
dimensional spacetime, have attracted a considerable amount of attention. Based on the braneworld model, all the particle fields in standard model are confined to a four-dimensional brane, whereas gravity is free to
propagate in all dimension. In particular, the proposed model by Dvali, Gabadadze
and Porrati (DGP) \cite{Dvali} is remarkably well suited to address the universe acceleration. In DGP braneworld model a 5D Minkowskian bulk induces
a late-time acceleration of the Universe expansion on the brane in agreement with current
observations. In addition, the energy–-momentum is confined on a
three-dimensional brane embedded in a five-dimensional, infinite volume
Minkowski bulk. According to the two ways that three dimensional brane embedded in five-dimensional bulk, the model features two separate branches denoted by $\epsilon$
with distinct characteristic.

Recently, a variety of attempts have been made to discuss dark energy models in DGP theory by adding a cosmological constant \cite{Sahni2}\cite{Lazkoz}, a Quintessence perfect fluid \cite{Chimento}, a scalar field \cite{Zhang} or a Chaplygin gas \cite{Lopez}. Alteratively, HDE in DGP models has also been investigated \cite{Wu}\cite{Liu}. This paper studies the evolution of the vacuum energy on the brane according to the agegraphic principle. The agegraphic dark energy model which relates the UV and IR cut-offs of a local quantum field system is well consistent with observational data. In comparison to HDE model \cite{Wu}, the ADE model predicts that the universe is in quintessence dominated regime, thus there is no super-acceleration and no big rip singularity in the model.

The paper is organized as follows: In section two, we formulate the original ADE in DGP braneworld model where "cosmic time scale" is taken to be the age of the universe. Alternatively, in section three, the age of the universe is assumed to be the "conformal time scale". Section four, is designed to constrain the model parameters in both cases. The observational probes are employed are Hubble parameter, cosmic microwave background anisotropies and baryon acoustic oscillation. We then study the dynamic of the DGP model in both original and new ADE scenarios.

\section{The original ADE in DGP model}

We start with the ADE density given by (\ref{rho}). In original ADE scenario, the cosmic time parameter, $t$, is chosen to be the age of the universe as
\begin{equation}\label{T}
T=\int_0^a\frac{da}{Ha}.
\end{equation}
Then, the energy density of the original ADE is in the form of \cite{Cai}
\begin{eqnarray}\label{rho}
\rho_{D}&=&\frac{3n^2M_p^2}{T^2},
\end{eqnarray}
where the numerical factor $3n^2$ parameterizes some uncertainties, such as the
species of quantum fields in the universe, the effect of curved space time and so on. Note that, the constant parameter $n$ plays the same role as $c$ in holographic dark energy
model. In the DGP braneworld cosmology, the universe is taken as a flat, homogeneous  and isotropic 3-dimensional brane embedded in 5-dimensional Minkowski bulk. The Friedmann
equation on the brane is \cite{Wu}
\begin{eqnarray}\label{fried}
H^2&=&(\sqrt{\frac{\rho}{3M_p^2}+\frac{1}{4r_c^2}}+\epsilon\frac{1}{2r_c})^2,
\end{eqnarray}
where $H=\dot a/a$ is the Hubble parameter and $r_c$ is called crossover distance.  As mentioned in teh introduction, the parameter $\epsilon=\pm1$ represents two branches of the model; the $\epsilon=+1$ branch is the self accelerating solution where the universe may accelerate in the late time purely due to modification of gravity, while the expansion of the $\epsilon=-1$ branch is not able to accelerate without dark energy. Here, to accommodate agegraphic dark energy into the formalism we take $\epsilon=-1$. we assume that the energy density $\rho$ in the brane contains both dust matter and dark energy, i.e., $\rho=\rho_m+\rho_D$. We proceed our model by introducing new dimensionless variables: the fractional energy densities such as
\begin{equation}\label{dimlesspar}
\Omega_{m}=\frac{\rho_{m}}{3M_p^2H^2},\quad\Omega_{D}=\frac{\rho_{D}}{3M_p^2H^2},\quad \Omega_{r_c}=\frac{1}{4r_c^2H_0^2}.
\end{equation}
These variables are indeed the fractional energy densities, where the Friedmann equation (\ref{fried}) can be rewritten in terms of them as
\begin{eqnarray}\label{fried2}
\Omega_{m}+\Omega_{D}+2\epsilon\frac{H_0}{H} \sqrt\Omega_{r_c}&=&1.
\end{eqnarray}
For later consideration we define a new dimensionless cosmological parameter related to the extra dimension as $\Omega_{DGP} = 2\epsilon\frac{H_0}{H}
\sqrt{\Omega_{r_c}}$, where the Friedmann equation (\ref{fried2}) in terms of new variable can be reexpressed as
\begin{equation}\label{fried3}
\Omega_m + \Omega_D +\Omega_{DGP} = 1.
\end{equation}
By using the energy density of the original ADE, (\ref{rho}), the new dynamical variable, $\Omega_{D}$, now becomes
\begin{eqnarray}\label{omega}
\Omega_{D}&=&\frac{n^2}{H^2T^2}.
\end{eqnarray}
By differentiating equation (\ref{rho}) with respect to the cosmic time and using equation (\ref{omega}) we obtain
\begin{eqnarray}\label{rhot}
\dot\rho_{D} = -2H\rho_D\frac{\sqrt{\Omega_D}}{n}.
\end{eqnarray}
From conservation equation for cold dark matter and dark energy in the brane we have
\begin{eqnarray}\label{ct1}
\dot{\rho_{m}}&+&3H\rho_{m}=0,
\end{eqnarray}
\begin{equation}\label{ct}
\dot{\rho_{D}} + 3H(1+\omega_{D})\rho_{D}=0.
\end{equation}
By inserting (\ref{rhot}) in the conservation equation (\ref{ct}) we find the EoS parameter of the original ADE as
\begin{eqnarray}\label{omega2}
\omega_{D}&=&-1+\frac{2}{3n}\sqrt{\Omega_D}.
\end{eqnarray}
Further, by differentiating (\ref{omega}) with respect to cosmic time $t$ and using $\dot\Omega_D=\Omega'_{D}H$, we also obtain
\begin{eqnarray}\label{omegap}
\Omega'_D &=& -2\Omega_D(\frac{\dot H}{H^2}+\frac{\sqrt{\Omega_D}}{n}),
\end{eqnarray}
where " $\prime$ " denotes derivative with respect to $x=\ln a$. From modified Friedmann equation (\ref{fried}) and conservation equations (\ref{ct1}) and (\ref{ct}), we find

\begin{eqnarray}\label{dH}
\frac{\dot H}{H^2} &=& -\frac{3}{2}(1-2\epsilon\frac{ H_0}{H}\sqrt{\Omega_{r_c}}-\Omega_D+\frac{2}{3n}\Omega_D^{3/2}) \nonumber\\ &\times&(1+\frac{H_0\epsilon\sqrt{\Omega_{r_c}}}{\sqrt{H_0^2\Omega_{r_c}
+H^2(1-2\epsilon\frac{H_0}{H}\sqrt{\Omega_{r_c}})}}).
\end{eqnarray}

Substituting the above relation into equation (\ref{omegap}), we obtain the following differential equation for the dimensionless density parameter of the original ADE
\begin{eqnarray}\label{omegap1}
\Omega'_D &=& \Omega_D(3(1-2\epsilon\frac{ H_0}{H}\sqrt{\Omega_{r_c}}-\Omega_D+\frac{2}{3n}\Omega_D^{3/2})\\&\times&(1+\frac{H_0\epsilon\sqrt{\Omega_{r_c}}}
{\sqrt{H_0^2\Omega_{r_c}+H^2(1-2\epsilon\frac{H_0}{H}\sqrt{\Omega_{r_c}})}})\nonumber
-\frac{2}{n}\sqrt{\Omega_D})
\end{eqnarray}
This equation describes the evolution of the original ADE in braneworld cosmology.
By using relation $\frac{d}{dx}=-(1+z)\frac{d}{dz}$ we can express $\Omega_D$ as
\begin{eqnarray}\label{omegaz}
\frac{d\Omega_D}{dz} &=& -(1+z)^{-1}\Omega_D(3(1-2\epsilon\frac{ H_0}{H}\sqrt{\Omega_{r_c}}-\Omega_D+\frac{2}{3n}\Omega_D^{3/2})\nonumber\\&\times&(1+\frac{H_0\epsilon\sqrt{\Omega_{r_c}}}
{\sqrt{H_0^2\Omega_{r_c}+H^2(1-2\epsilon\frac{H_0}{H}\sqrt{\Omega_{r_c}})}})
-\frac{2}{n}\sqrt{\Omega_D}).
\end{eqnarray}
Also by using equation (\ref{dH}) we can write
\begin{eqnarray}\label{dH2}
\frac{dH}{dz} &=& \frac{3}{2}H(1+z)^{-1}(1-2\epsilon\frac{ H_0}{H}\sqrt{\Omega_{r_c}}-\Omega_D+\frac{2}{3n}\Omega_D^{3/2})\nonumber\\&\times&(1+\frac{H_0\epsilon\sqrt{\Omega_{r_c}}}
{\sqrt{H_0^2\Omega_{r_c}+H^2(1-2\epsilon\frac{H_0}{H}\sqrt{\Omega_{r_c}})}}).
\end{eqnarray}
It is difficult to solve equations (\ref{omegaz}) and (\ref{dH2}) analytically, however, the evolutionary form of the equation of motion of agegraphic dark energy in DGP cosmology and
Hubble parameter can be obtained by integrating them numerically from $z = 0$ to a given value
$z$. In addition, from relation (\ref{dH}), one can easily find the deceleration parameter as

\begin{eqnarray}\label{q2}
 q&=&-1 +\frac{3}{2}(1-2\epsilon\frac{ H_0}{H}\sqrt{\Omega_{r_c}}-\Omega_D+\frac{2}{3n}\Omega_D^{3/2}) \nonumber\\ &\times&(1+\frac{H_0\epsilon\sqrt{\Omega_{r_c}}}
 {\sqrt{H_0^2\Omega_{r_c}+H^2(1-2\epsilon\frac{H_0}{H}\sqrt{\Omega_{r_c}})}}).
\end{eqnarray}

\section{The new ADE in DGP Model}

The original ADE model contain some internal inconsistencies such as difficulty to describe the matter-dominated epoch \cite{Wei}. New ADE, due to estimating a good approximate of dark energy value and solving causality problem of HDE, have
received huge interest. In this section we propose the so-called new agegraphic dark energy with some new features that overcome unsatisfactory points in the original ADE. In new ADE model the length scale is chosen to be the conformal age $\eta$ instead of the cosmic age of the universe and given by
\begin{equation}\label{eta}
\eta=a_0\int_0^a \frac{da}{Ha^2}.
\end{equation}
The energy density of the new ADE can then be written as
\begin{eqnarray}\label{rhon}
\rho_{D}&=&\frac{3n^2M_p^2}{\eta^2}.
\end{eqnarray}
From dimensionless dynamical variables, the fractional energy density of the new ADE is given by
\begin{eqnarray}\label{omegan}
\Omega_{D}&=&\frac{n^2}{H^2\eta^2}
\end{eqnarray}
Taking derivative with respect to the cosmic time of equation (\ref{rhon}) and using equation (\ref{omegan}) we get
\begin{eqnarray}\label{rhotn}
\dot\rho_{D} = -2H\rho_D\frac{a_0\sqrt{\Omega_D}}{na}.
\end{eqnarray}
Inserting this equation in the conservation law (\ref{ct}) we can find the EoS parameter for new ADE as
\begin{eqnarray}\label{omega2n}
\omega_{D}&=&-1+\frac{2a_0}{3na}\sqrt{\Omega_D}.
\end{eqnarray}
We can
also find equation of motion for $\Omega_D$ as
\begin{eqnarray}\label{omegazn}
\frac{d\Omega_D}{dz} &=& -(1+z)^{-1}\Omega_D(3(1-2\epsilon\frac{ H_0}{H}\sqrt{\Omega_{r_c}}-\Omega_D+\frac{2a_0}{3na}\Omega_D^{3/2})\nonumber\\&\times&(1+\frac{H_0\epsilon\sqrt{\Omega_{r_c}}}
{\sqrt{H_0^2\Omega_{r_c}+H^2(1-2\epsilon\frac{H_0}{H}\sqrt{\Omega_{r_c}})}})-\frac{2a_0}{na}\sqrt{\Omega_D})
\end{eqnarray}
Similar to the original ADE scenario, we can easily find the differential equation for Hubble parameter as
\begin{eqnarray}\label{dH2n}
\frac{dH}{dz} &=& \frac{3}{2}H(1+z)^{-1}(1-2\epsilon\frac{ H_0}{H}\sqrt{\Omega_{r_c}}-\Omega_D+\frac{2a_0}{3na}\Omega_D^{3/2})\nonumber\\&\times&(1+\frac{H_0\epsilon\sqrt{\Omega_{r_c}}}
{\sqrt{H_0^2\Omega_{r_c}+H^2(1-2\epsilon\frac{H_0}{H}\sqrt{\Omega_{r_c}})}}).
\end{eqnarray}
Solving the set of differential equations (\ref{dH2n}) and (\ref{omegazn}), numerically, we find the solutions for $H(z)$ and $\Omega_D$. Also deceleration parameter can be found as

\begin{eqnarray}\label{q2n}
 q&=&-1 +\frac{3}{2}(1-2\epsilon\frac{ H_0}{H}\sqrt{\Omega_{r_c}}-\Omega_D+\frac{2a_0}{3na}\Omega_D^{3/2}) \nonumber\\ &\times&(1+\frac{H_0\epsilon\sqrt{\Omega_{r_c}}}
 {\sqrt{H_0^2\Omega_{r_c}+H^2(1-2\epsilon\frac{H_0}{H}\sqrt{\Omega_{r_c}})}}).
\end{eqnarray}

In the next section, we constraint the model parameters in both original and new ADE of DGP model with observational data for Hubble parameter, BAO distance ratio and the cosmic microwave background (CMB).

\section{Cosmological constraints}

We study the constraints on the model parameters using $\chi^2$ method, for recent observational data of Hub, BAO distance ratio and CMB radiation. The $\chi^2$ for Hubble parameter is:
\begin{eqnarray}\label{chi2}
    &&\chi^2_{Hub}( n ,\Omega_{rc},H_0;\Omega_{D}(0))=\nonumber\\
   && \sum_{i=1}^{14}\frac{[H^{the}(z_i|n ,\Omega_{rc},H_0;\Omega_{D}(0)) - H^{obs}(z_i)]^2}{\sigma_{Hub}^2(z_i)},
\end{eqnarray}
where the sum is over the cosmological data points. In (\ref{chi2}), $H^{the}$ and $H^{obs}$ are the Hubble parameters obtained from theoretical model and observation, respectively. Also, $\sigma_{Hub}$ is the estimated error of the $H^{obs}$ obtained from observation.

\begin{figure}[h]
\centering
\includegraphics[width=0.3\textwidth]{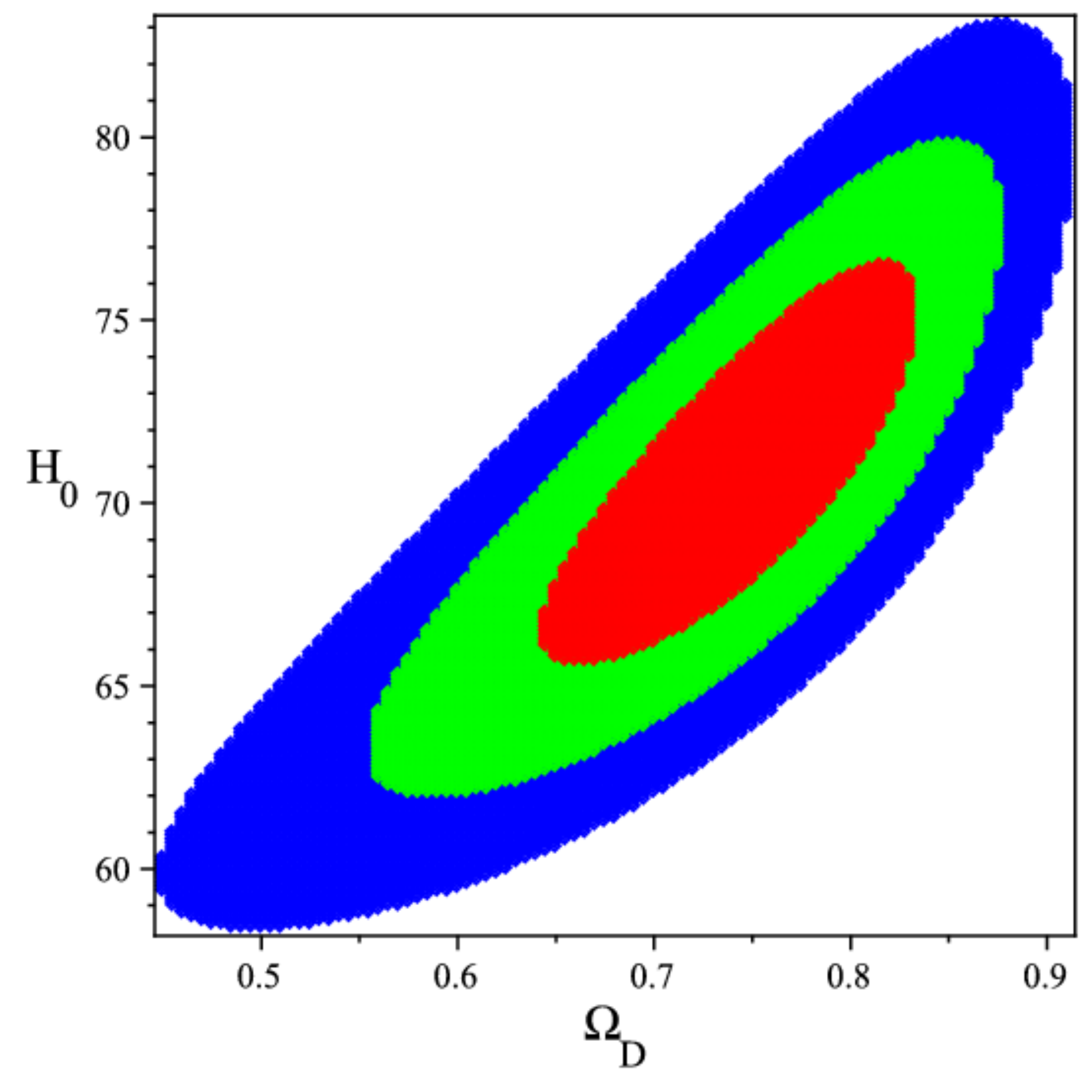}
\includegraphics[width=0.3\textwidth]{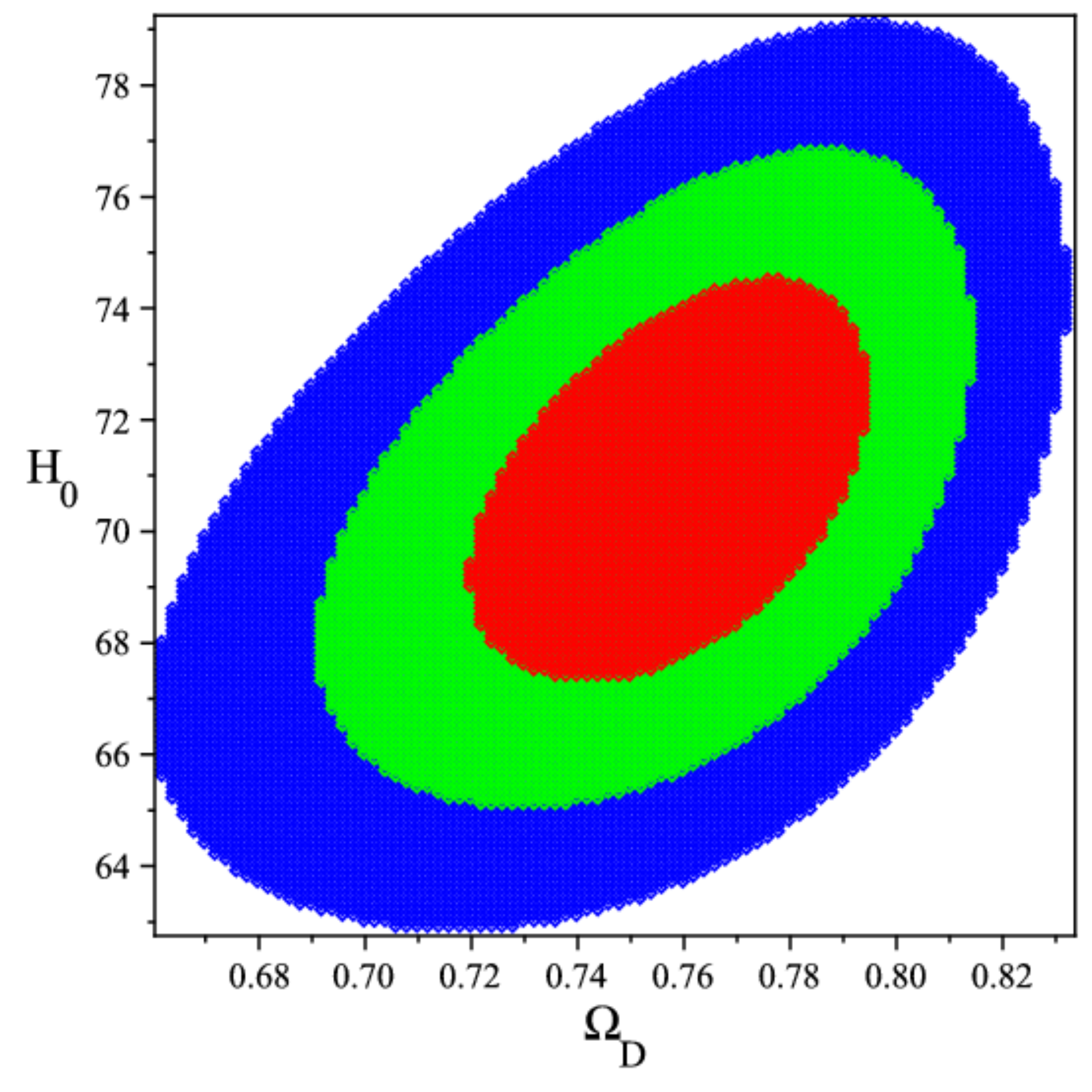}
\includegraphics[width=0.3\textwidth]{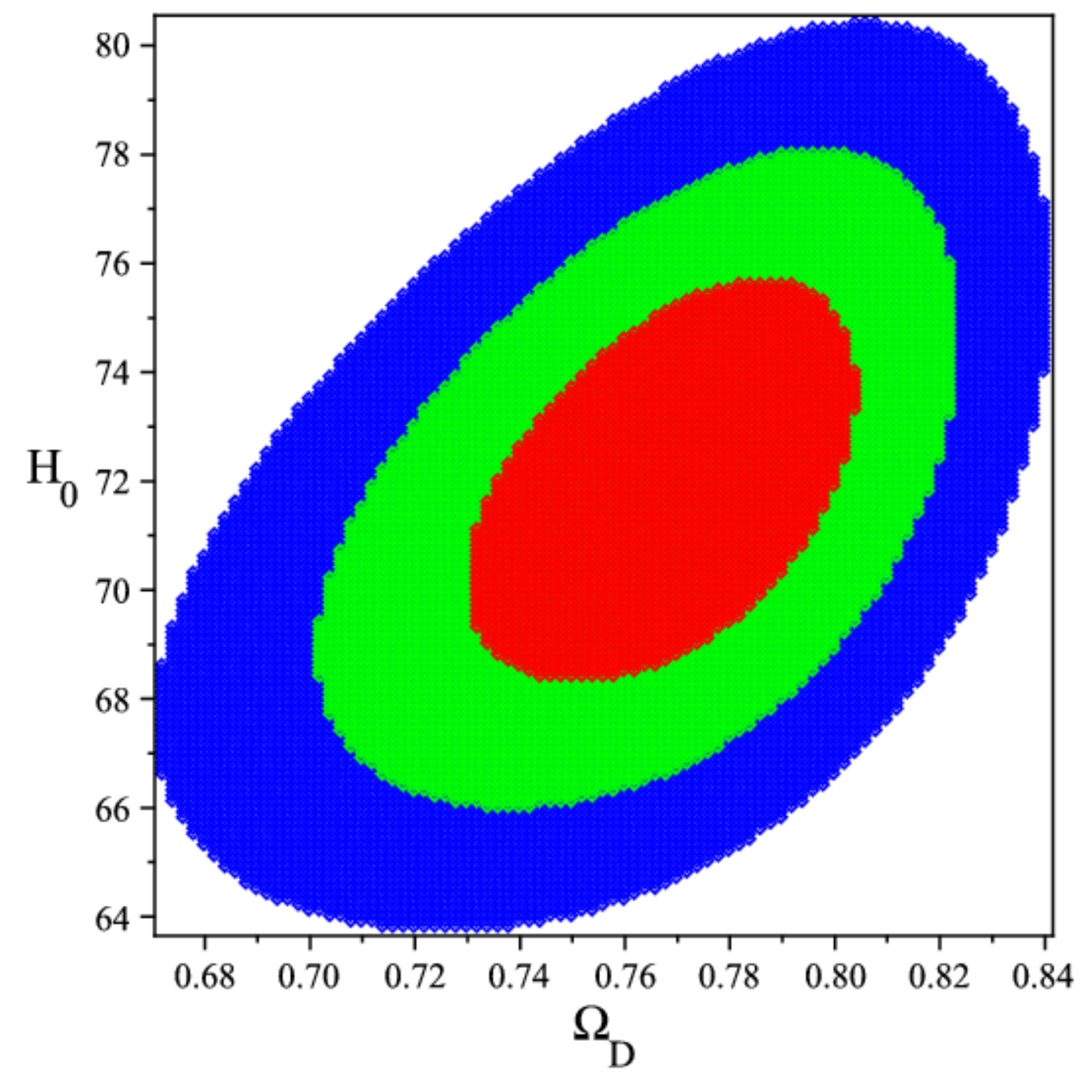}
\caption{The constraints on $\Omega_{r_c}$ and $\Omega_{D0}$ at $1\sigma$ (red), $2\sigma$ (green) and $3\sigma$ (blue) confidence regions from H, H+CMB and H+CMB+BAO (from left to right, respectively) in original ADE model.}\label{fig1}
\end{figure}

\begin{figure}[t]
\centering
\includegraphics[width=0.3\textwidth]{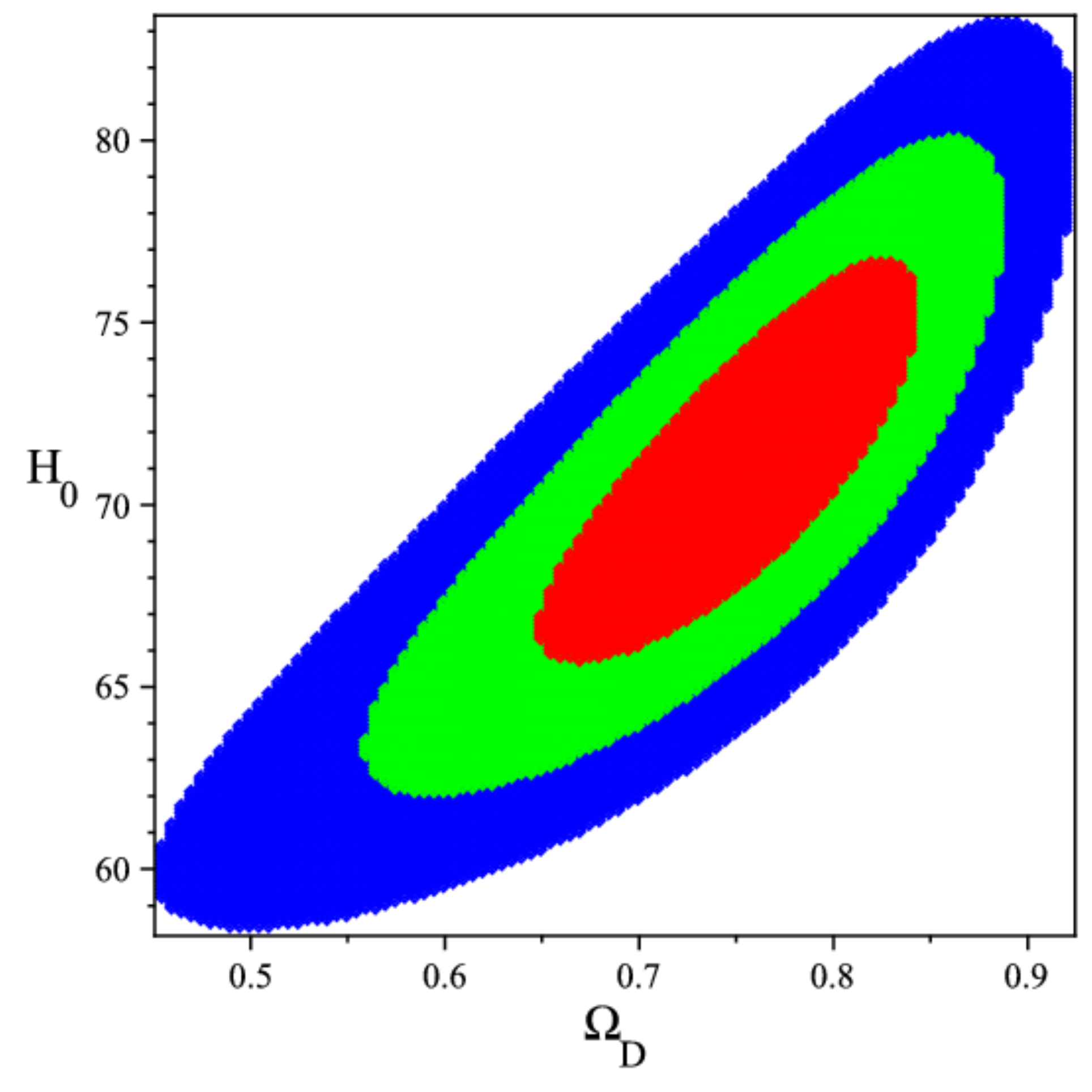}
\includegraphics[width=0.3\textwidth]{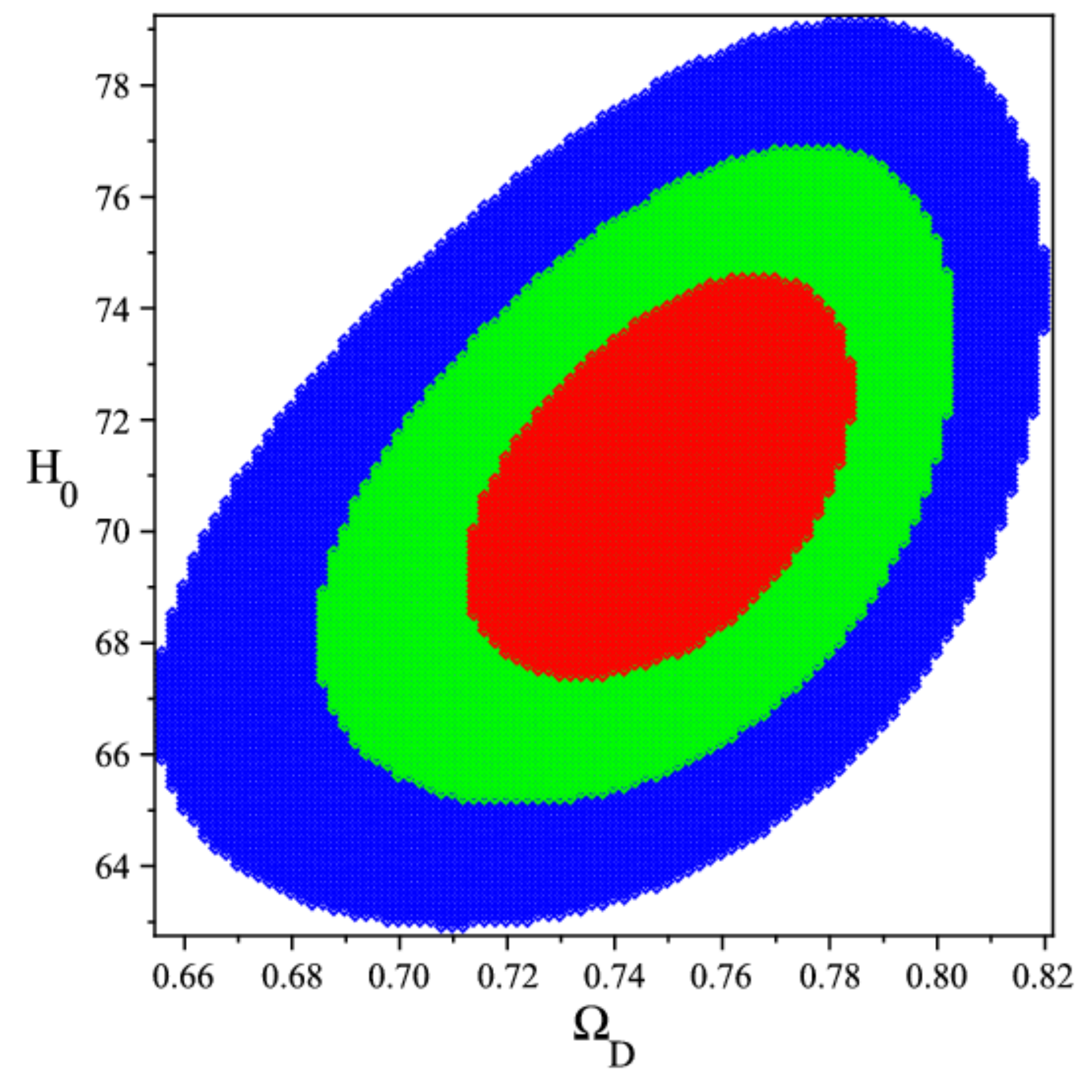}
\includegraphics[width=0.3\textwidth]{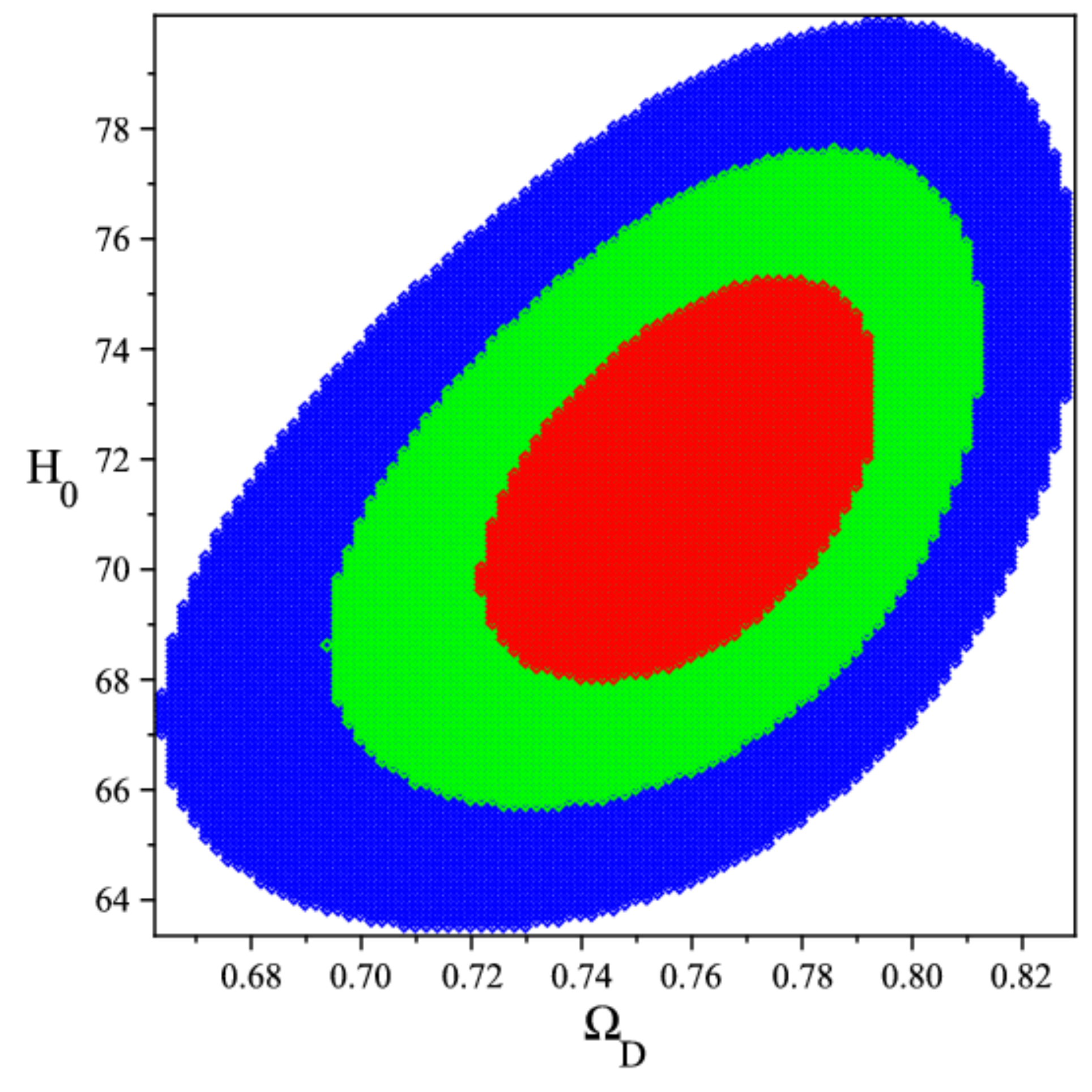}
\caption{The constraints on $\Omega_{r_c}$ and $\Omega_{D0}$ at $1\sigma$ (red), $2\sigma$ (green) and $3\sigma$ (blue) confidence regions from H, H+CMB and H+CMB+BAO (from left to right, respectively) in NADE model.}\label{fig2}
\end{figure}

In addition, the CMB shift parameter $R$, contains the main information of the observations from CMB, thus, it can be used to constrain the theoretical models by minimizing  \cite{Wang}\cite{Bond}
\begin{equation}\label{chicmb}
    \chi^2_{CMB}=\frac{[R-R_{obs}]^2}{\sigma_R^2},
\end{equation}
where $R_{obs} = 1.725\pm0.018$ \cite{Komatsu1}, is given by $WMAP7$ data. The CMB shift parameter, $R$, is defined as
\begin{equation}\label{r}
    R\equiv\Omega_{m0}^{1/2}\int_0^{z_{CMB}}\frac{dz'}{E(z')},
\end{equation}
with $z_{CMB} = 1091.3$. Furthermore, at $z = 0.20$ and $z = 0.35$ by using the BAO distance ratio from joint analysis of the 2dF Galaxy Redsihft Survey and SDSS data \cite{Reid}\cite{Reid1} we can constrain the model parameters. The BAO distance defined as
\begin{equation}\label{bao}
D_V(z_{BAO})=[\frac{z_{BAO}}{H(z_{BAO})}(\int_0^{z_{BAO}}\frac{dz}{H(z)})^2]^{1/3}.
\end{equation}
The distance ratio at $z = 0.20$ and $z = 0.35$
\begin{equation}\label{drbao}
   \frac{D_V(z=0.35)}{D_V(z=0.20)}=1.736\pm0.065,
\end{equation}
is a model independent quantity. As a result, the constraint from BAO can be obtained by performing the following $\chi^2$ test,
\begin{equation}\label{chibao}
    \chi^2_{BAO}=\frac{[(D_V(z=0.35)/D_V(z=0.20))-1.736]^2}{0.065^2}\cdot
\end{equation}
From a combination of Hub,  BAO and CMB and by minimizing $\chi^2_{Hub}+\chi^2_{BAO}+\chi^2_{CMB}$ one can find the constraints. The best fitted parameters, $H_0$, $\Omega_{D0}$, $\Omega_{r_c}$ and $n$ are shown in Tables \ref{table:1} and \ref{table:2} for two original and new ADE scenarios. Also, the contour plots for confidence levels of the model parameters have been shown in figures \ref{fig1} and \ref{fig2}. From Tables I and II, we see that the best fitted values of the parameters, $H_0$, $\Omega_{D0}$, and $\Omega_{r_c}$ in both scenarios are the same, whereas constraint on parameter $n$ is model dependent. The best fitted values in the Tables, indicate that DGP cosmological model with respect to original and new ADE scenarios does not alter with respect to the first three model parameters, while changes with $n$. This can easily be justified by comparing equations (\ref{dH2n}) and (\ref{q2n}) with  equations (\ref{dH2}) and (\ref{q2}).  Thus, one expects that the cosmological behaviors in two scenarios are affected by only the presence of parameter $n$ in the equations.

\begin{table}[ht]
\caption{bestfit values of original ADE model} 
\centering 
\begin{tabular}{|c|c|c|c|} 
\hline\hline 
observational data  &  H \ & H+CMB \ & H+CMB+BAO \\ [4ex] 
\hline 
$H_0$ & $71^{+2}_{-2}$ & $71^{+2}_{-2}$ & $72^{+2}_{-2}$ \\ 
\hline 
$\Omega_{D0}$ & $0.75^{+0.03}_{-0.03}$ & $0.76^{+0.02}_{-0.02}$ & $0.77^{+0.02}_{-0.02}$ \\
\hline 
$\Omega_{r_c}$& $0.0003^{+0.0011}_{-0.0003}$ & $0.0003^{+0.0004}_{-0.0002}$ & $0.0003^{+0.0004}_{-0.0002}$\\
\hline
n & $18^{+\infty}_{-14}$ & $7^{+\infty}_{-3}$ & $17^{+\infty}_{-12}$\\
\hline
\end{tabular}
\label{table:1} 
\end{table}\

\begin{table}[ht]
\caption{bestfit values of NADE model} 
\centering 
\begin{tabular}{|c|c|c|c|} 
\hline\hline 
observational data  &  H \ & H+CMB \ & H+CMB+BAO \\ [4ex] 
\hline 
$H_0$ & $71^{+2}_{-2}$ & $71^{+2}_{-2}$ & $72^{+2}_{-2}$ \\ 
\hline 
$\Omega_{D0}$ & $0.75^{+0.03}_{-0.03}$ & $0.76^{+0.02}_{-0.02}$ & $0.77^{+0.02}_{-0.02}$ \\
\hline
$\Omega_{r_c}$& $0.0003^{+0.0011}_{-0.0003}$ & $0.0003^{+0.0004}_{-0.0002}$ & $0.0003^{+0.0004}_{-0.0002}$\\
\hline 
n & $21^{+\infty}_{-16}$ & $10^{+\infty}_{-5}$ & $21^{+\infty}_{-14}$\\
\hline
\end{tabular}
\label{table:2} 
\end{table}\

\section{Cosmological parameters}

Using the best-fitted constrained model parameters, we now discuss the behavior of EoS parameter in ADE and NADE models, deceleration parameter and dimensionless cosmological parameters in DGP braneworld model.

\begin{figure}[t]
\centering
\includegraphics[width=0.3\textwidth]{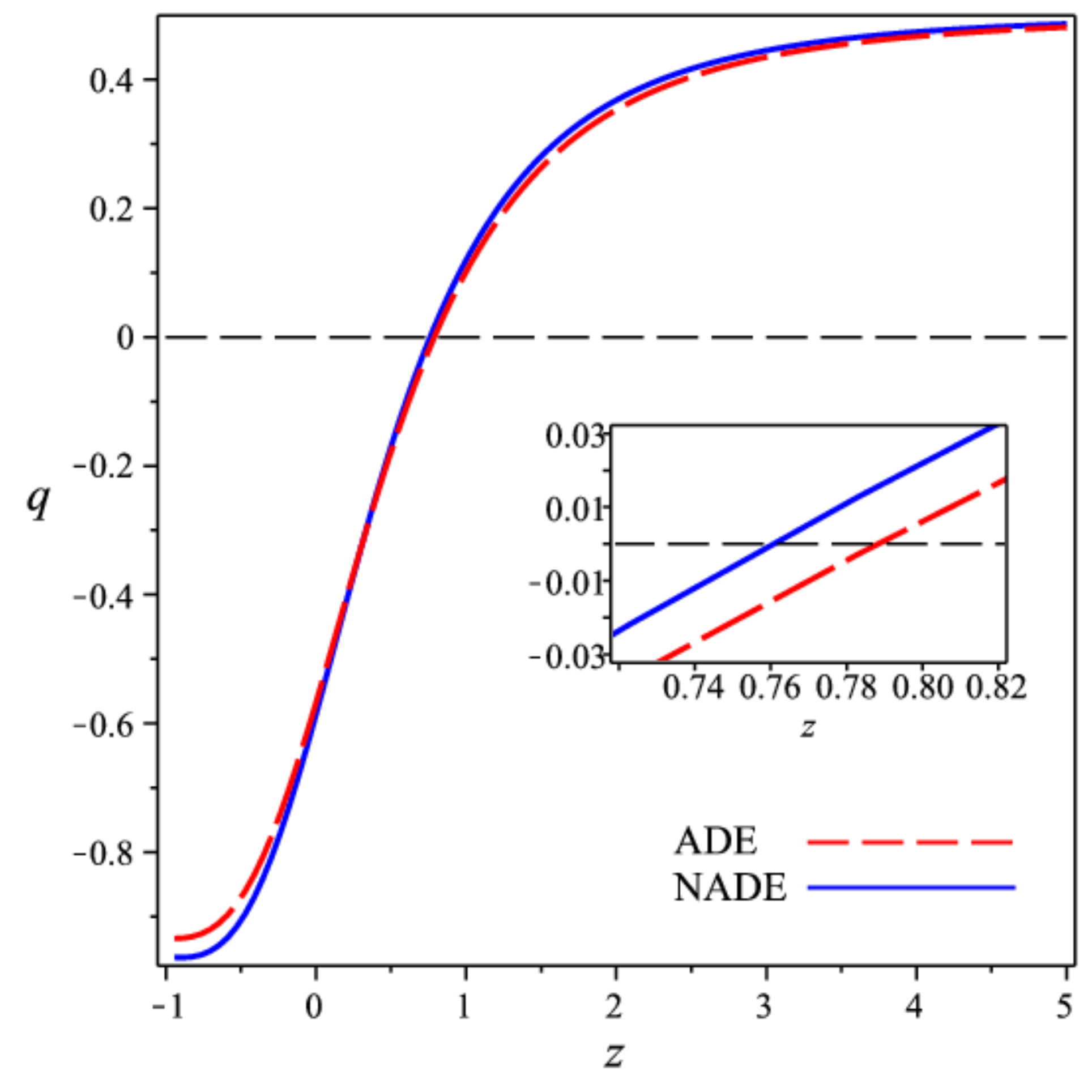}
\includegraphics[width=0.3\textwidth]{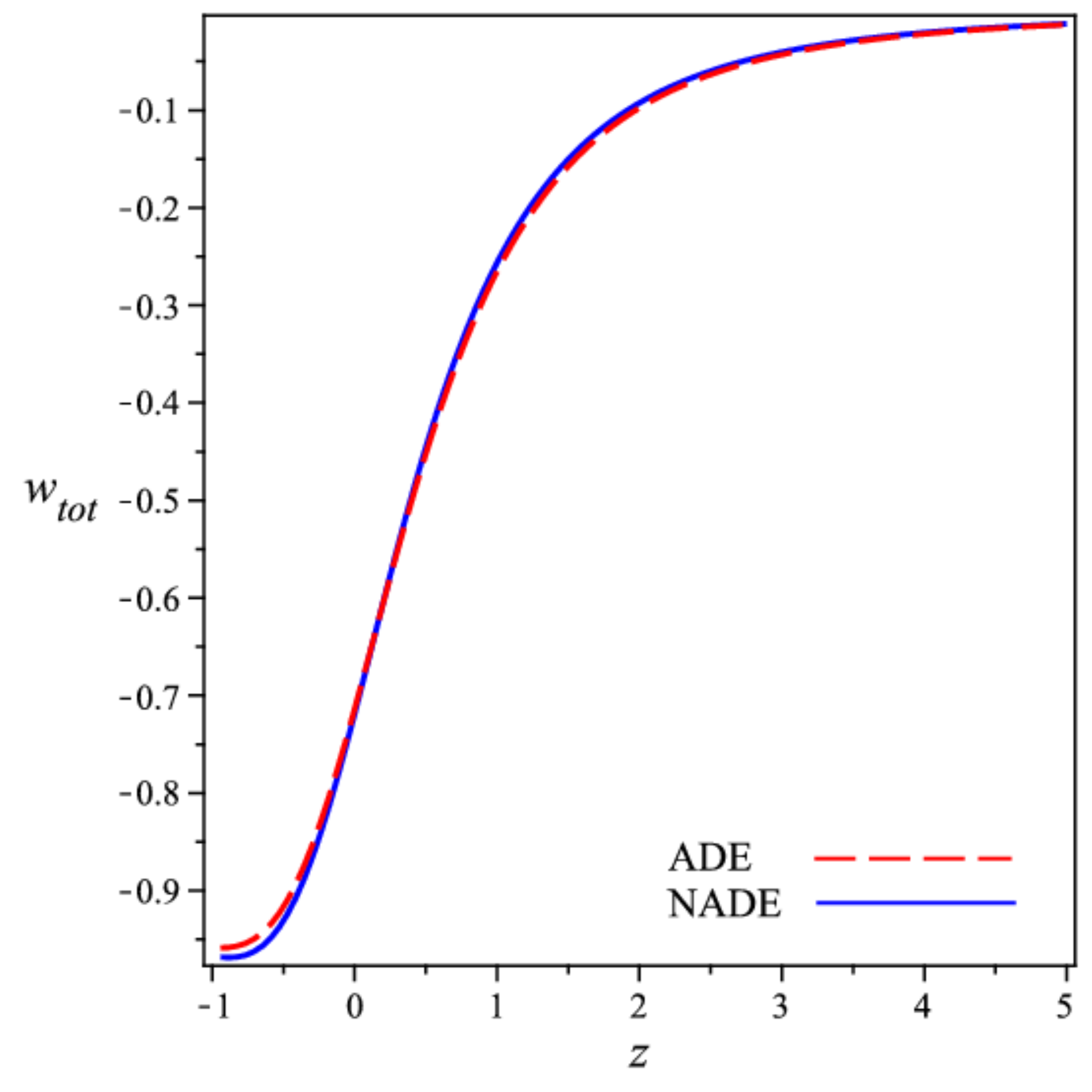}
\includegraphics[width=0.3\textwidth]{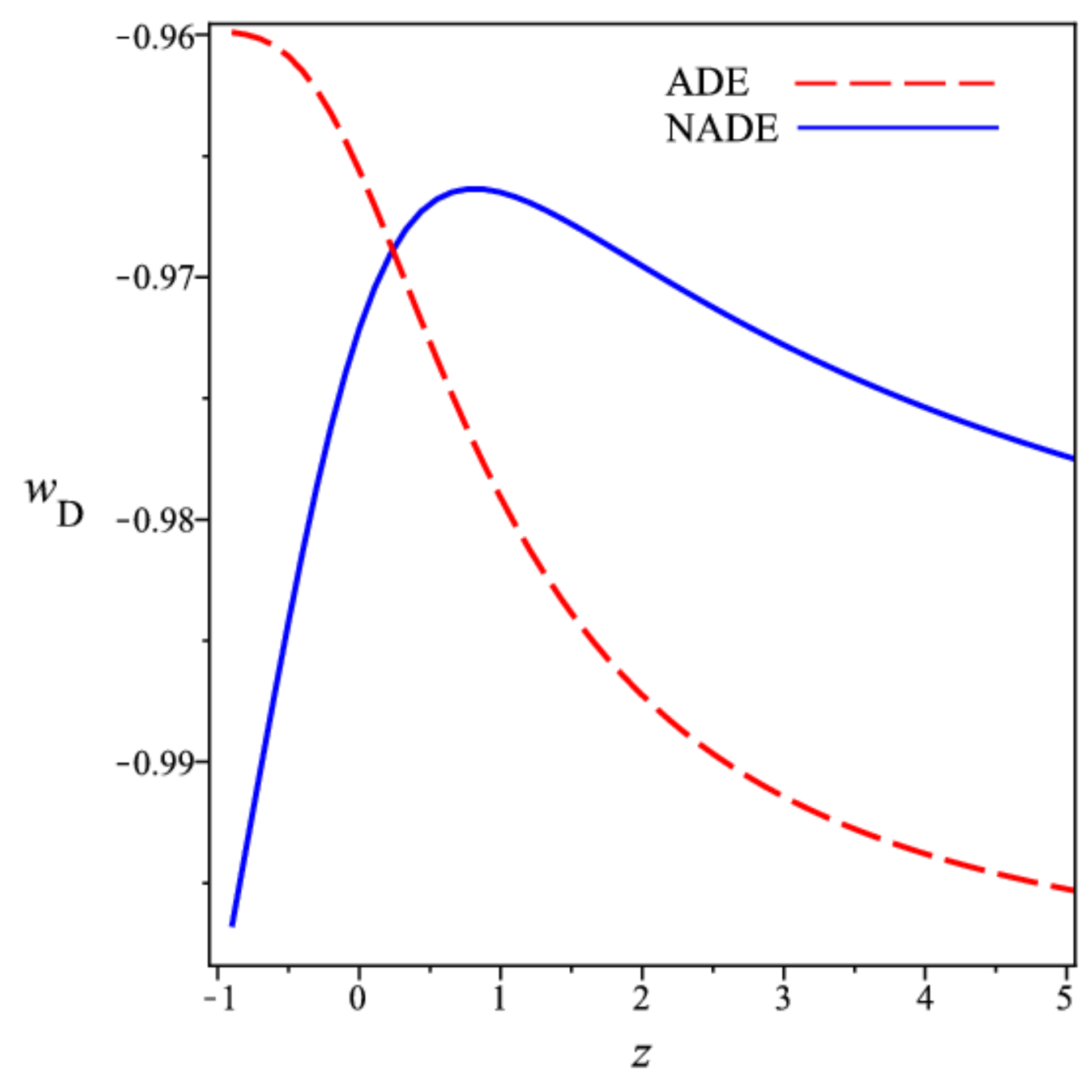}
\caption{left panel: The evolutionary curves of the deceleration parameter for the best fitted values of combining H+CMB+BAO. middle panel: The evolutionary curves of the total EoS parameter for the best fitted values of combining H+CMB+BAO. left panel: The evolutionary curves of dark energy EoS parameter for the best fitted values of combining H+CMB+BAO. The dash red curve and the solid blue curve show the original ADE and the NADE models, respectively.}\label{fig3}
\end{figure}

In figure \ref{fig3} left panel, we show the dynamic of the dark energy EoS parameter of the original and new ADE models for the best fit parameters. Also, figure \ref{fig3} right panel illustrates the behavior of deceleration parameter in both cases. From the graph, in the original ADE model for dark eenrgy, the universe begins to accelerate  at $z \sim 0.79$, whereas in NADE case at $z \sim 0.76$. The behavior of the deceleration parameter in both scenarios are very much the same; only in recent epoch they slightly deviate from each other. we obtain q=-0.5 for the present value of the deceleration parameter which is in good agreement with recent observational results.
 For EoS parameter, though they show different behavior in near past up to around $z\simeq 1$, in higher redshifts, both EoS parameters similarly decrease with increasing redshift. The middle panel in figure \ref{fig3} shows the total EoS parameter in both original and new ADE cases. The graph shows in both cases the behavior of total EoS parameter of the model is the same. Further, the cosmic undergoes acceleration without entering phantom regime in the past or future as predicted in HDE model of DGP cosmology.

\begin{figure}[t]
\centering
\includegraphics[width=0.3\textwidth]{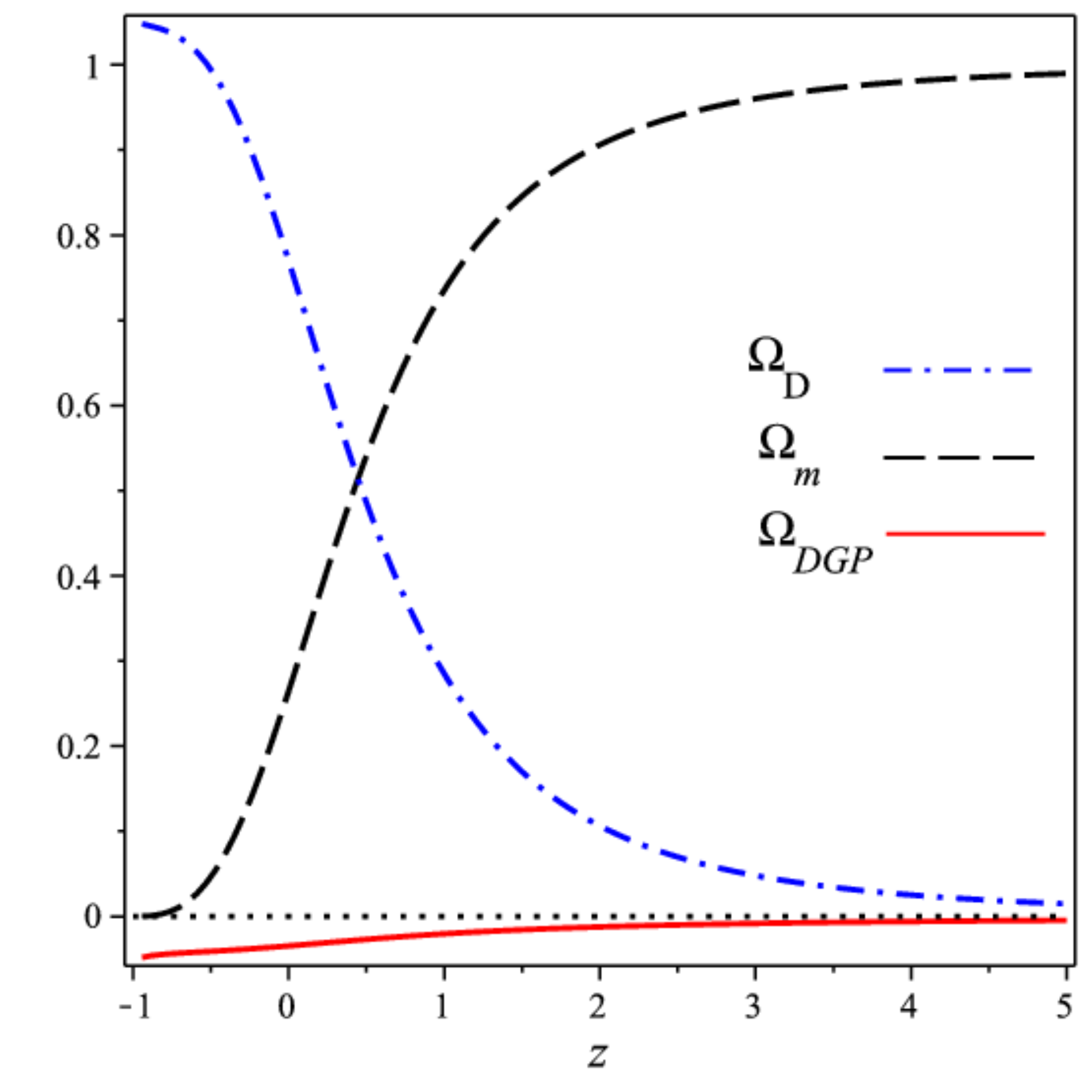}
\includegraphics[width=0.3\textwidth]{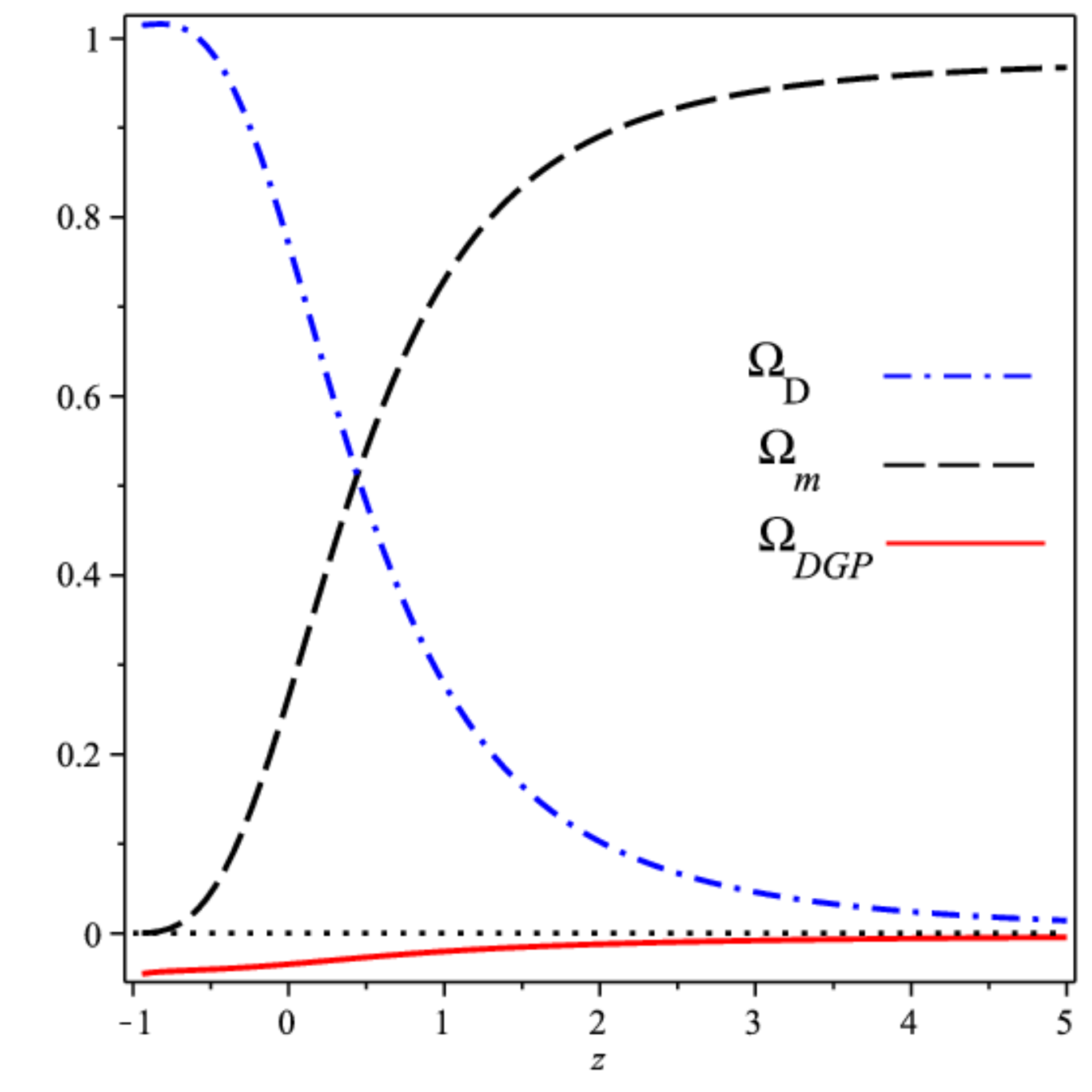}
\caption{The evolutionary curves of the dimensionless density parameters $\Omega_m$ (black dash curve), $\Omega_D$ (blue dash-dot curve) and $\Omega_{DGP}$ (solid red curve) for the best fitted values of of combining H+CMB+BAO in both cases original ADE model (left) and NADE model (right).}\label{fig4}
\end{figure}

On the other hand, the dynamic of dimensionless density parameters, $\Omega_m$, $\Omega_D$ and $\Omega_{DGP}$, expression (\ref{dimlesspar}), are shown in \ref{fig4}. As expected for $\epsilon=-1$, the DPG density parameter, $\Omega_{DGP}$, makes a negative dynamical contribution to the energy constraint equation (\ref{fried3}). However, although its effect on the dynamic of the total density parameter diminishes in higher redshift in both original and new ADE scenarios, in low redshift it has a greater effect in new ADE scenario than original one. From figure \ref{fig4}, we also see that the current value of density parameter for cold dark matter in the universe is about $0.27$

In summary, this work analysis the original and new ADE scenarios in DGP cosmology. We first constrained the model parameters with the observational data for Hub, BAO and CMB, using $\chi^2$ method. We found that the values of the three best fitted parameters and initial conditions, $\Omega_{rc}$, $H_0$ and $\Omega_{D}(0)$ are the same in both original and new ADE cases whereas the best fitted constant parameter $n$ is different. The analysis of the dynamical equations in two scenarios also implies and verifies the numerical finding. The DGP model in both scenarios confirms current universe acceleration in quintessence era. From the graph of energy constraint in both original and new ADE (figure 4) we see that the negative value of DGP density parameter plays as a compensating dynamical term into the constraint equation. The graph also shows that while the universe is in matter dominated phase in the past, both DGP and ADE (original and new) density parameter vanish.

\end{document}